\documentclass[aps,prb,reprint,preprintnumbers,superscriptaddress,amsmath,amssymb,bibnotes,longbibliography]{revtex4-2}
\usepackage{graphicx}
\usepackage{dcolumn}
\usepackage{epstopdf}
\usepackage{braket}
\usepackage{bm,multirow,threeparttable}
\usepackage{flushend}
\usepackage[pdfstartview=FitH, CJKbookmarks=true, bookmarksnumbered=true, bookmarksopen=true, colorlinks, pdfborder=001, linkcolor=blue, anchorcolor=blue, citecolor=blue,urlcolor=blue]{hyperref}
\newcommand{\red}[1]{\textcolor{black}{#1}}

\begin{document}
\title{Magnetic properties of  $R$Rh$_6$Ge$_4$ ($R$ = Pr, Nd, Sm, Gd - Er) single crystals}
\author{Jiawen Zhang}
\affiliation  {Center for Correlated Matter and School of Physics, Zhejiang University, Hangzhou 310058, China}
\author{Yongjun Zhang}
\affiliation  {Institute for Advanced Materials, Hubei Normal University, Huangshi 435002, China}
\author{Yuxin Chen}
\affiliation  {Center for Correlated Matter and School of Physics, Zhejiang University, Hangzhou 310058, China}
\author{Zhaoyang Shan}
\affiliation  {Center for Correlated Matter and School of Physics, Zhejiang University, Hangzhou 310058, China}
\author{Jin Zhan}
\affiliation  {Center for Correlated Matter and School of Physics, Zhejiang University, Hangzhou 310058, China}
\author{Mingyi Wang}
\affiliation  {Center for Correlated Matter and School of Physics, Zhejiang University, Hangzhou 310058, China}

\author{Yu Liu}
\affiliation  {Center for Correlated Matter and School of Physics, Zhejiang University, Hangzhou 310058, China}
\author{Michael Smidman}
\email[Corresponding author: ]{msmidman@zju.edu.cn}
\affiliation  {Center for Correlated Matter and School of Physics, Zhejiang University, Hangzhou 310058, China}
\author{Huiqiu Yuan}
\email[Corresponding author: ]{hqyuan@zju.edu.cn}
\affiliation  {Center for Correlated Matter and School of Physics, Zhejiang University, Hangzhou 310058, China}
\affiliation  {Institute for Advanced Study in Physics, Zhejiang University, Hangzhou 310058, China}
\affiliation  {Institute of Fundamental and Transdisciplinary Research, Zhejiang University, Hangzhou 310058, China}
\affiliation  {State Key Laboratory of Silicon and Advanced Semiconductor Materials, Zhejiang University, Hangzhou 310058, China}

\date{\today}

\begin{abstract}
Single crystals of $R$Rh$_6$Ge$_4$ ($R$ = Pr, Nd, Sm, Gd - Er) were synthesized using a Bi flux and their physical properties were characterized by magnetization, resistivity, and specific heat measurements. These compounds crystallize in the noncentrosymmetric LiCo$_6$P$_4$-type structure (space group $P\bar{6}m2$), where rare-earth atoms form a triangular lattice in the $ab$-plane and chains along the $c$-axis. PrRh$_6$Ge$_4$ and ErRh$_6$Ge$_4$ do not exhibit magnetic transitions above 0.4 K. NdRh$_6$Ge$_4$ and SmRh$_6$Ge$_4$ are ferromagnets, while GdRh$_6$Ge$_4$ and DyRh$_6$Ge$_4$ show antiferromagnetic transitions, \red{whereas HoRh$_6$Ge$_4$ is a ferrimagnet}. In addition, DyRh$_6$Ge$_4$ shows multiple transitions and magnetization plateaus when a magnetic field is applied along the $c$-axis. In SmRh$_6$Ge$_4$, like the Ce counterpart, the crystalline-electric field (CEF) effect leads to an easy plane anisotropy, while in other compounds it gives rise to a pronounced uniaxial anisotropy.

\begin{description}
\item[PACS number(s)]
\end{description}
\end{abstract}

\maketitle

\section{Introduction}
\red{Lanthanide compounds show rich physical properties, including superconductivity \cite{Pfl2009,Smidan2023}, quantum phase transitions \cite{Si2010}, strange metal behavior \cite{ShenB2020,phillips2022stranger}, and topological electronic structures \cite{Wenlong2021}. Their ground states can be tuned via non-thermal parameters, such as pressure, magnetic field \cite{Kne2006}, chemical doping \cite{Weng_2016}.}
The intermetallic materials $R$Rh$_6X_4$ ($R$ = rare earth, $X$ = Si, Ge) crystallize in the hexagonal LiCo$_6$P$_4$-type structure \cite{VosswinkelD2012,VosswinkelD2013,MatsuokaE2015} shown in Fig. \ref{figure1}(a) and (b), in which relatively closely spaced rare-earth atoms make up chains running along the $c$-axis, that form a triangular lattice in the basal plane. 
CeRh$_6$Ge$_4$ is a ferromagnetic (FM) Kondo lattice with $T_{\rm C}$ = 2.5 K \cite{ShenB2020,MatsuokaE2015}, which was found to exhibit a FM quantum critical point (QCP) upon tuning with hydrostatic pressure \cite{ShenB2020,Kot2019}, near which there is strange metal behavior \cite{ShenB2020} characterized by a $T$-linear resistivity and a logarithmic divergence of the specific heat coefficient.
Although the electronic structure of CeRh$_6$Ge$_4$ is three-dimensional \cite{WangA2021}, a pronounced anisotropy in the $c$-$f$ hybridization is observed in angle-resolved photoemission spectroscopy (ARPES) \cite{WuY2021}, which may reflect a dominant Ruderman-Kittel-Kasuya-Yosida (RKKY) interaction along the Ce chains. Theoretical studies indicate that such a quasi-one-dimensional (q1D) magnetic anisotropy could be key for the observed FM QCP \cite{ShenB2020,Komi2018}. A crystalline-electric field (CEF)-level scheme with a large positive value of $B_2^0$ is proposed to account for the easy-plane magnetocrystalline anisotropy and the anisotropy of the CEF orbitals is considered important for the anisotropic $c$-$f$ hybridization \cite{PhysRevB.104.L140411}.

Considering the apparent intricate role of the $f$-electron states in leading to the unusual behaviors, it is of particular interest to study $R$Rh$_6$Ge$_4$ with other lanthanide elements, which modifies the 4$f$ electron count.
Previously, antiferromagnetic (AFM) transitions have been observed at 8.4 K, 13.6 K, 5.1 K, and 8.9 K for $R$ = Gd, Tb, Dy, and Yb, respectively \cite{VosswinkelD2013}. 
We recently synthesized high quality single crystals of TbRh$_6$Ge$_4$ and characterized the anisotropic magnetic properties \cite{ChenY2023}. In TbRh$_6$Ge$_4$, we observed an additional transition at 2.8 K and a significant anisotropy with multiple metamagnetic transitions along the $c$-axis, where there are plateaus in the field-dependence of the magnetization at $1/9$ and $1/3$ of the saturation value $M_s$ \cite{ChenY2023}. The field-temperature phase diagram reveals complex magnetism in TbRh$_6$Ge$_4$, in which magnetic frustration may play an important role \cite{ChenY2023}.

In this study, we report the single crystal growth and characterization of the anisotropic magnetic properties of $R$Rh$_6$Ge$_4$ ($R$ = Pr, Nd, Sm, Gd - Er).  For $R$ = Nd, Sm, Dy and Ho, we observed a significant magnetocrystalline anisotropy due to the CEF effects. 
\red{Two new ferromagnets NdRh$_6$Ge$_4$ and SmRh$_6$Ge$_4$ are identified with $T_{\rm C}$ at 2.26 K and 1.65 K, respectively. We found the magnetization plateau at $M_s$/3 in ferrimagnet HoRh$_6$Ge$_4$ with $T_{\rm M} \sim 2.28$ K, and observed an additional transition at 1.7 K as well as magnetization plateaus at $M_s$/9 and $M_s$/3 in DyRh$_6$Ge$_4$. Finally, we constructed the field-temperature ($H-T$) phase diagram of DyRh$_6$Ge$_4$.}

\begin{table}[!ht]
	\renewcommand\arraystretch{1.4}
	\caption{Lattice parameters, unit cell volume, and $c/a$ ratio for $R$Rh$_6$Ge$_4$.}
	\begin{tabular}{lllll}
		\hline\hline
		$R$~~~~~~ & $a$ (\AA)~~~~~~~~~~ & $c$ (\AA)~~~~~~~~~~ & Volume (\AA$^3$)~~~~~~ & $c/a$~~~~ 
		\\ \hline
		Pr & 7.1454(2) & 3.8462(2) & 170.065(13) & 0.5383
		\\
		Nd & 7.1436(3) & 3.8368(2) & 169.564(17) & 0.5371
		\\
		Sm & 7.1406(1) & 3.8227(1) & 168.799(6) & 0.5353
		\\
		Gd & 7.101(4) & 3.798(3) & 165.9(2) & 0.5349 
		\\
		Tb & 7.127(4) & 3.800(3) & 167.2(2) & 0.5332 
		\\
		Dy & 7.147(3) & 3.802(2) & 168.19(13) & 0.5320 
		\\
		Ho & 7.1380(2) & 3.7885(2) & 167.167(13) & 0.5308
		\\
		Er & 7.1469(11) & 3.7856(10) & 167.46(7) & 0.5297
		\\ \hline
	\end{tabular}
	\label{table1}
\end{table}

\section{Experimental methods}

Single crystals of $R$Rh$_6$Ge$_4$ ($R$ = Pr, Nd, Sm, Gd - Er) were grown using a Bi-flux method \cite{VosswinkelD2013,ShenB2020}. The constituent elements were placed in an alumina crucible in a molar ratio $R$ : Rh : Ge : Bi = 1 : 6 : 4 : 100 and sealed in an evacuated quartz ampule. The ampule was heated to 1100 $^\circ$C over 10 h, held at this temperature for 10 h, and then cooled to 500 $^\circ$C at a rate of 2 $^\circ$C/h. Finally, needle-like shiny single crystals were obtained by centrifuging the excess flux. Residual Bi flux on the surface of single crystals could be dissolved in a 1:1 molar mixture of H$_2$O$_2$ and acetic acid.

The chemical composition was checked by using energy-dispersive x-ray analysis with a Hitachi SU-8010 field emission scanning electron microscope, which confirms the atomic ratio to be 1 : 6 : 4. The crystal structure was characterized using single-crystal x-ray diffraction (XRD) with a Bruker D8 Venture diffractometer with Mo K$_\alpha$ radiation. The magnetization and magnetic susceptibility measurements were performed using a Quantum Design Magnetic Property Measurement System (MPMS-5T) \red{from 1.8 K to 300 K} and a MPMS-VSM that is equipped with a $^3$He refrigerator \red{from 0.4 K to 1.8 K}. The specific heat was measured down to 0.4 K using the relaxation method in a Quantum Design Physical Property Measurement System (PPMS), and the electrical resistivity was measured using a standard four-probe technique in a PPMS. 

\section{Results and discussion}

\subsection{A. Crystal structure}

\begin{figure}[htbp]
	\begin{center}
		\includegraphics[width=\columnwidth]{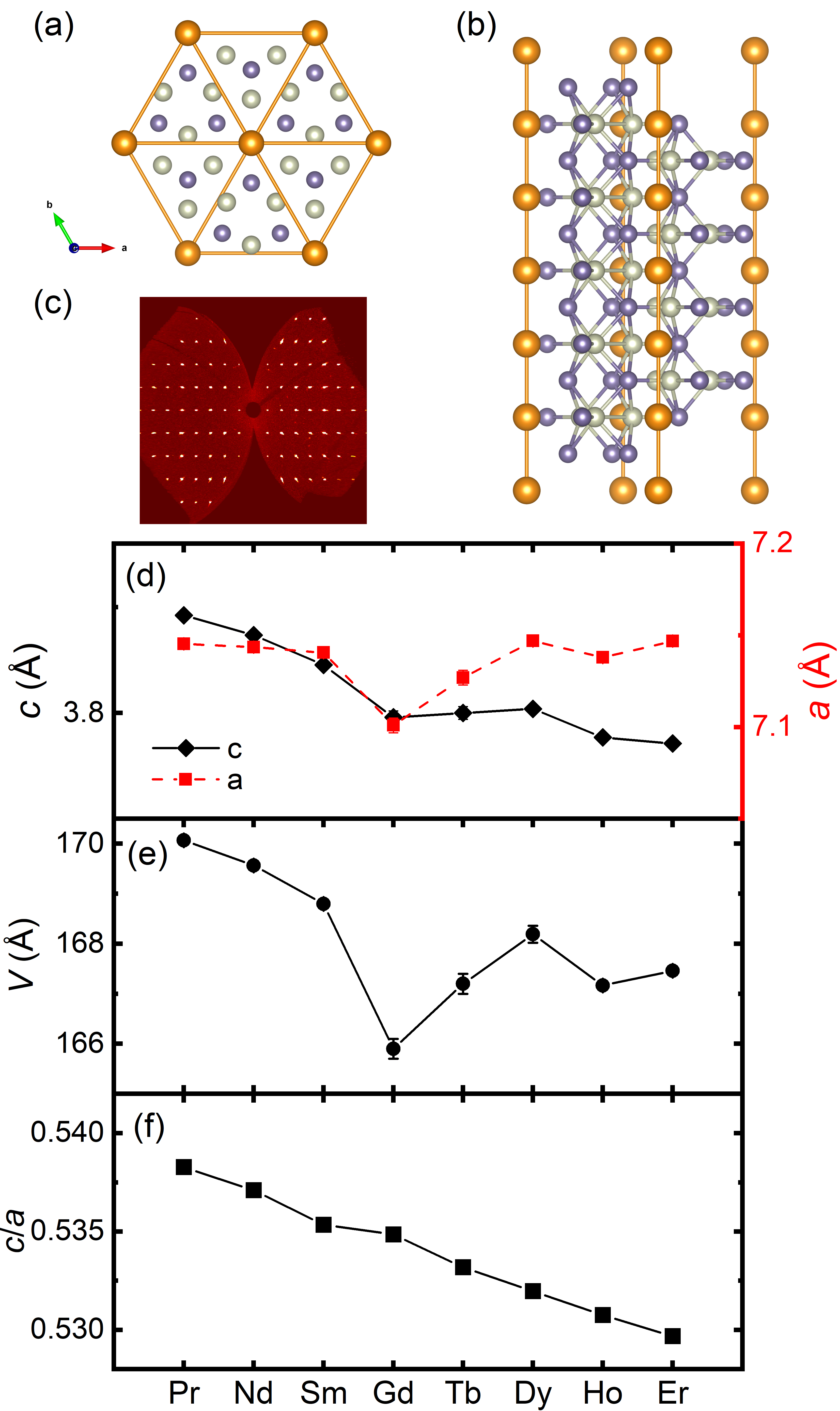}
	\end{center}
	\caption{(Color online) Crystal structure of $R$Rh$_6$Ge$_4$ ($R$ = Pr, Nd, Sm, Gd - Er), viewed (a) parallel, and (b) perpendicular to the chain direction ($c$ axis), where orange, gray, and blue represent $R$, Rh, and Ge atoms, respectively. (c) Single crystal XRD pattern of a GdRh$_6$Ge$_4$ crystal in the (0 h l) plane. (d) Lattice constants, (e) unit cell volume, and (f) $c/a$ ratio for $R$Rh$_6$Ge$_4$.}
	\label{figure1}
\end{figure}

Single crystal XRD was performed so as to characterize the crystal structure. 
All the $R$Rh$_6$Ge$_4$ ($R$ = Pr, Nd, Sm, Gd - Er) compounds crystallize in the hexagonal LiCo$_6$P$_4$-type structure ($P\bar{6}m2$), consistent with the previous report \cite{VosswinkelD2013}, and the obtained lattice parameters are listed in Table \ref{table1}. As shown in Fig. \ref{figure1}(c), a representative single crystal XRD pattern of GdRh$_6$Ge$_4$ indicates high quality structurally ordered single crystals. 

The obtained lattice constants $a$ and $c$ of $R$Rh$_6$Ge$_4$ are plotted in Fig. \ref{figure1}(d). Besides Gd and Tb, the value of $c$ shrinks gradually from Pr to Er while $a$ is almost unchanged. As shown in Fig. \ref{figure1}(e), the change of the unit cell volume is largely consistent with the lanthanide contraction, besides Gd and Tb. However, as depicted in Fig. \ref{figure1}(f), the $c/a$ value decreases monotonically from Pr to Er, suggesting an increasingly q1D arrangement of the rare-earth atoms.

\subsection{B. Paramagnetic compounds: PrRh$_6$Ge$_4$ and ErRh$_6$Ge$_4$}

\begin{figure}[htbp]
	\begin{center}
		\includegraphics[width=0.95\columnwidth]{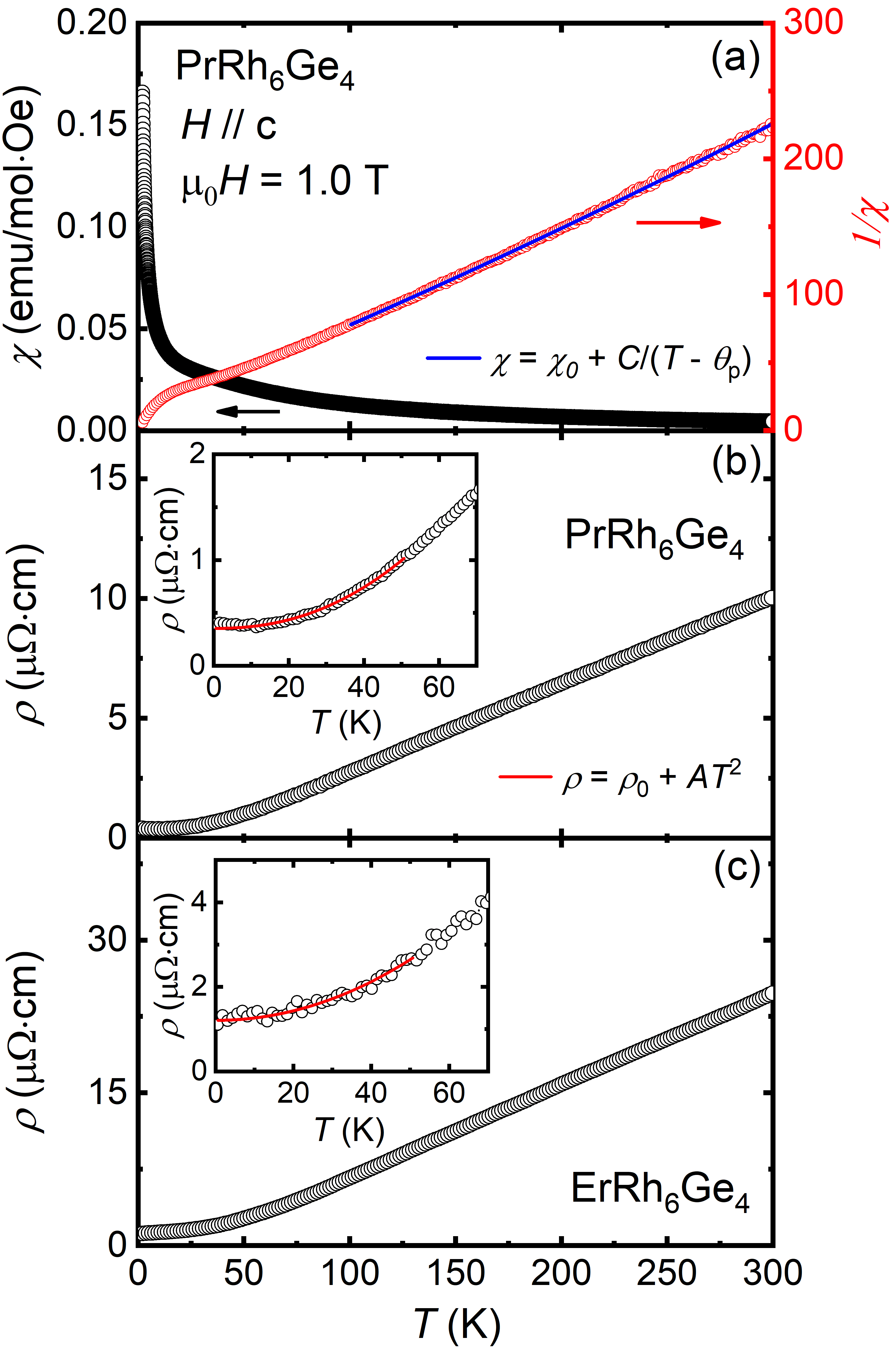}
	\end{center}
	\caption{(Color online) (a) \red{Temperature dependence of the magnetic susceptibility $\chi(\rm T)$ (left axis) and 1/$\chi$ (right axis) of PrRh$_6$Ge$_4$ measured in $\mu_0H$ = 1 T for $H$ $\Vert$ $c$, where the solid line corresponds to fit to the modified Curie-Weiss law}. Temperature dependence of the resistivity $\rho(\rm T)$ of (b) PrRh$_6$Ge$_4$ and (c) ErRh$_6$Ge$_4$. The insets show $\rho(\rm T)$ at low temperatures. Red lines show fitting using $\rho = \rho_0 + AT^2$.}
	\label{figure2}
\end{figure}

PrRh$_6$Ge$_4$ and ErRh$_6$Ge$_4$ do not exhibit magnetic transitions down to 0.4 K. As shown in Fig. \ref{figure2}(a), the magnetic susceptibility $\chi(\rm T)$ of PrRh$_6$Ge$_4$ follows paramagnetic behaviors above 2.0 K. \red{PrRh$_6$Ge$_4$ follows a modified Curie-Weiss behavior, $\chi({\rm T}) = \chi_0 + C/(T-\theta_{\rm p})$, at high temperatures, where $\chi_0$ is the temperature-independent term. An effective magnetic moment of $\mu_{{\rm eff}}$ = 3.56 $\mu_{\rm B}$/Pr, a Curie-Weiss temperature $\theta_{\rm p}$ = -18.2(3) K and $\chi_0$ = -0.00057(1) emu/mol$\cdot$Oe were obtained by fitting the $1/\chi$ curves above 100 K to the modified Curie-Weiss law. The estimated effective moment of PrRh$_6$Ge$_4$ is close to the free ion value of 3.5 $\mu_{\rm B}$/Pr for Pr$^{3+}$.}

The temperature dependence of the resistivity $\rho(\rm T)$ measured with the current parallel to the $c$-axis (Fig. \ref{figure2}(b) and (c)), follows typical metallic behavior without any transitions down to 0.4 K. The residual resistivity ratio (RRR) of PrRh$_6$Ge$_4$ and ErRh$_6$Ge$_4$ are 26 and 21, respectively. 
\red{Below 50 K, $\rho(\rm T)$ of both compounds can be fitted by $\rho = \rho_0 + AT^2$, where the second term corresponds to the Fermi liquid contribution. The fitted parameters are $\rho_0$ = 0.351(4) $\mu\Omega\cdot$cm and $A$ = 0.000249(4) $\mu\Omega\cdot$cm$\cdot$K$^{-2}$ for PrRh$_6$Ge$_4$, while $\rho_0$ = 1.23(2) $\mu\Omega\cdot$cm and $A$ = 0.00055(2) $\mu\Omega\cdot$cm$\cdot$K$^{-2}$ for ErRh$_6$Ge$_4$.}


\subsection{C. Ferromagnetic compounds: NdRh$_6$Ge$_4$ and SmRh$_6$Ge$_4$}

\begin{figure}[htbp]
	\begin{center}
		\includegraphics[width=0.95\columnwidth]{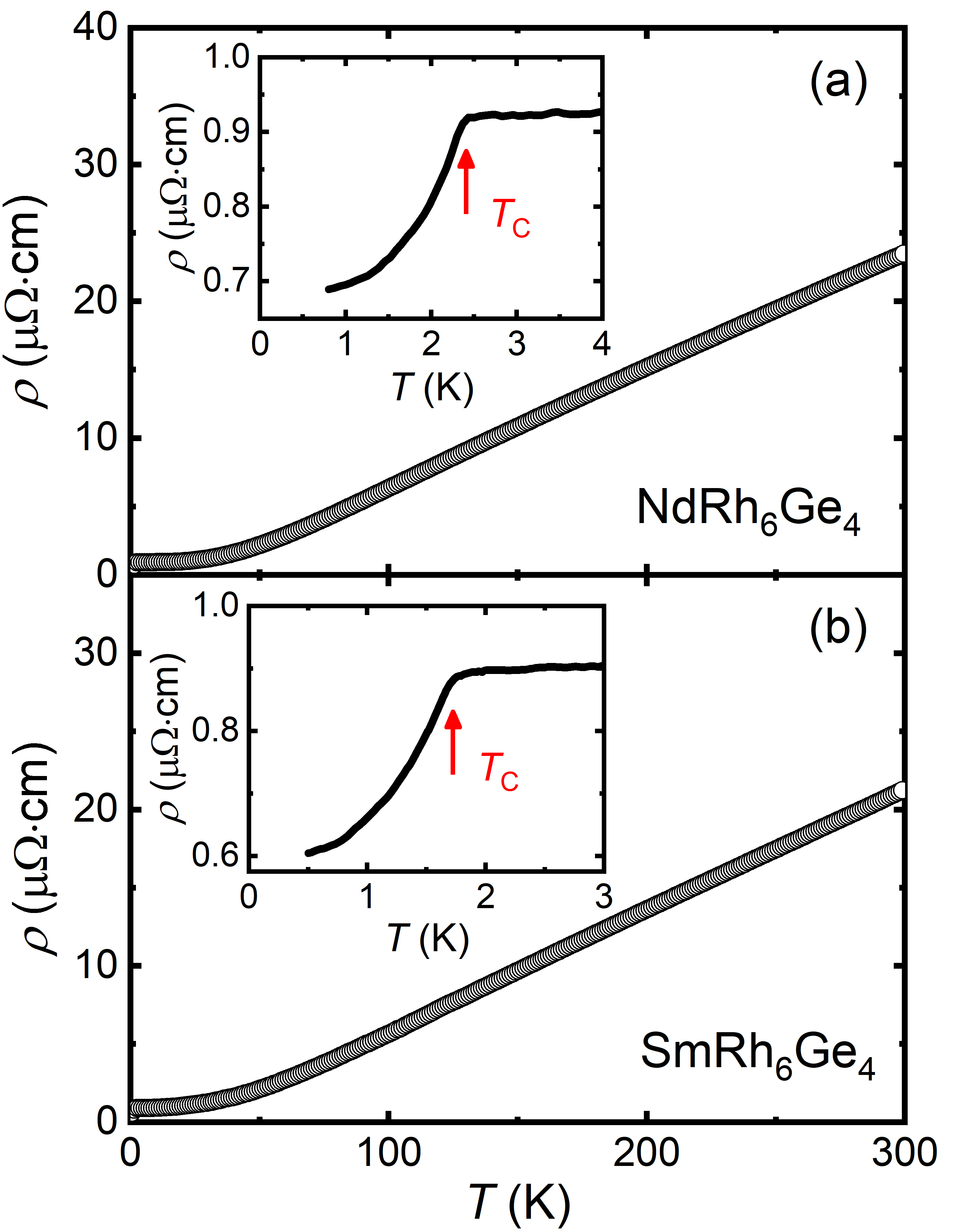}
	\end{center}
	\caption{(Color online) Temperature dependence of the resistivity $\rho(\rm T)$ of (a) NdRh$_6$Ge$_4$ and (b) SmRh$_6$Ge$_4$. Insets show $\rho(\rm T)$ at low temperatures, where red arrows indicate the magnetic transitions.}
	\label{figure3}
\end{figure}

The $\rho(\rm T)$ of NdRh$_6$Ge$_4$ and SmRh$_6$Ge$_4$ are displayed in Fig. \ref{figure3}(a) and (b), which show a clear drop at 2.26 K and 1.65 K, respectively.

\begin{figure*}[htbp]
	\begin{center}
		\includegraphics[width=0.85\textwidth]{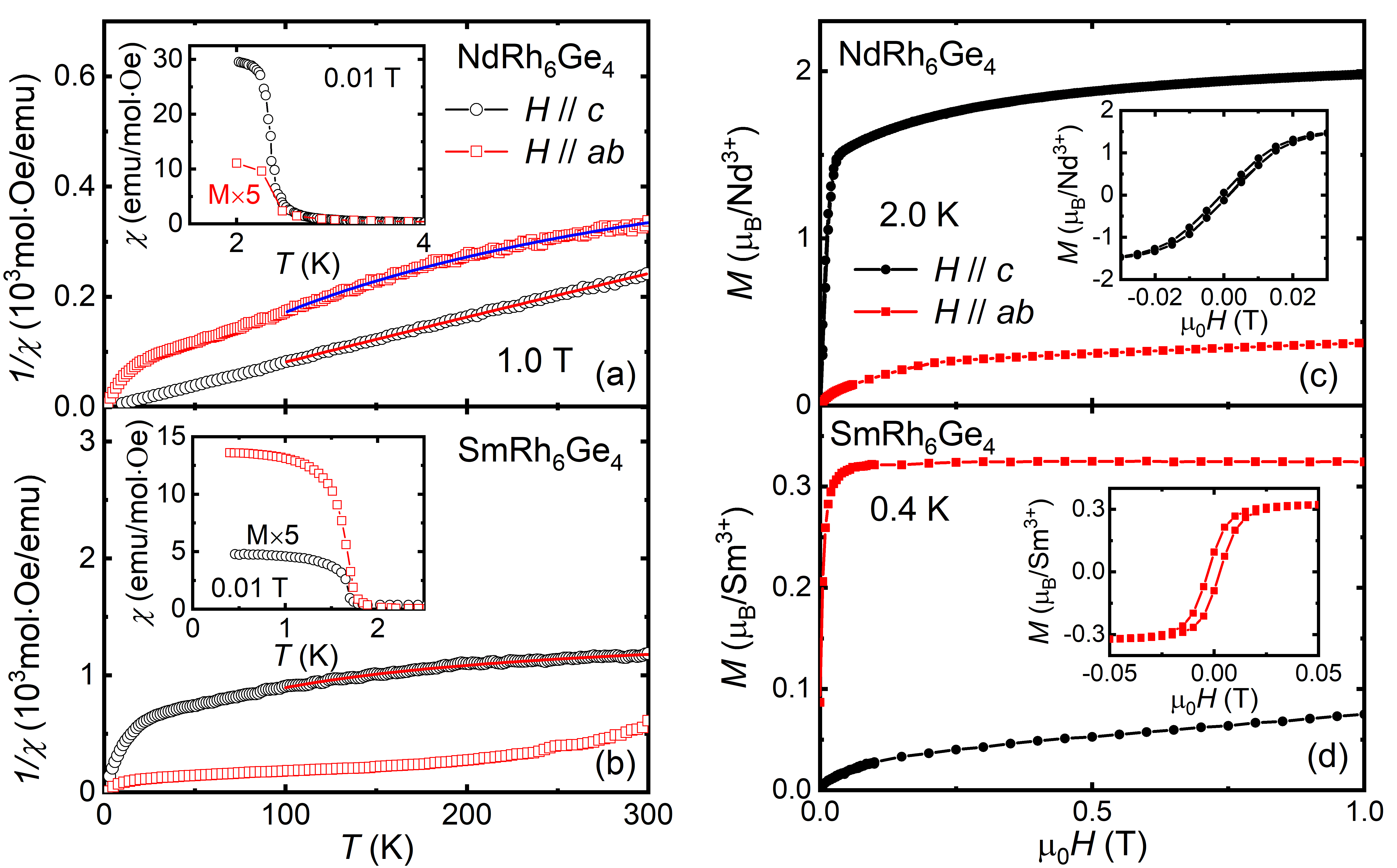}
	\end{center}
	\caption{(Color online) (a) and (b) show the respective temperature dependence of the inverse magnetic susceptibility $1/\chi$ measured in $\mu_0H$ = 1 T for $H$ $\Vert$ $c$ (circles) and $H$ $\Vert$ $ab$ (squares), where the solid lines correspond to fits to the modified Curie-Weiss law. Insets show the magnetic susceptibility $\chi(\rm T)$ \red{measured upon field-cooling (FC)} at low temperatures. (c) and (d) show the field dependence of the magnetization for $H$ $\Vert$ $c$ (circles) and $H$ $\Vert$ $ab$ (squares). The insets show the magnetization loops with hysteresis in the low-field region.}
	\label{figure4}
\end{figure*}

To determine the nature of these transitions, we measured magnetic susceptibility for both compounds for $H$ $\Vert$ $c$ and $H$ $\Vert$ $ab$ with $\mu_0H$ = 1 T (Fig. \ref{figure4}(a) and (b)). NdRh$_6$Ge$_4$ follows a modified Curie-Weiss behavior at high temperatures.
For $H$ $\Vert$ $c$, an effective magnetic moment of $\mu_{{\rm eff}}$ = 3.05 $\mu_{\rm B}$/Nd, a Curie-Weiss temperature $\theta_{\rm p}$ = 3.3(9) K and $\chi_0$ = 0.00021(4) emu/mol$\cdot$Oe were obtained by fitting the $1/\chi$ curves above 150 K to the modified Curie-Weiss law, while for $H$ $\Vert$ $ab$, $\mu_{{\rm eff}}$ = 2.97 $\mu_{\rm B}$/Nd, $\theta_{\rm p}$ = -18(3) K and $\chi_0$ = -0.00834(9) emu/mol$\cdot$Oe. The estimated effective moments of NdRh$_6$Ge$_4$ are smaller than the free ion value of 3.62 $\mu_{\rm B}$/Nd for Nd$^{3+}$, which may be due to the influence of CEF effects. 
For SmRh$_6$Ge$_4$, the $1/\chi$ within the $ab$-plane can be analyzed using a modified Curie-Weiss
law, giving a negative $\theta_{\rm p}$ of -19(7) K and $\chi_0$ = 0.00069(1) emu/mol$\cdot$Oe. The estimated effective moment of 0.64 $\mu_B$/Sm is somewhat smaller than the theoretical Sm$^{3+}$ free ion value of $\mu_{{\rm eff}}$ = 0.84 $\mu_B$/Sm. For the $c$-axis, it does not conform to the modified Curie-Weiss law, which is common in Sm-containing compounds with mixed-valent Sm-ions \cite{Mazumda1994,NapGM1978}.
The insets of Fig. \ref{figure4}(a) and (b) show the magnetic susceptibility $\chi(\rm T)$ of NdRh$_6$Ge$_4$ and SmRh$_6$Ge$_4$ at low temperatures. There are sharp increases around the magnetic transitions, and $\chi(\rm T)$ eventually becomes saturated, which are typical FM behaviors.

The field dependence of the magnetization, $M(\rm H)$, at 2.0 K for NdRh$_6$Ge$_4$ and 0.4 K for SmRh$_6$Ge$_4$ show a large anisotropy between $H$ $\Vert$ $c$ and $H$ $\Vert$ $ab$, as shown in Fig. \ref{figure4}(c) and (d). The easy magnetization direction is along the $c$-axis for NdRh$_6$Ge$_4$ and in the $ab$-plane for SmRh$_6$Ge$_4$, \red{the difference of which is likely related to the single-ion magnetocrystalline anisotropy arising from the influence of CEF.}
The magnetization of NdRh$_6$Ge$_4$ and SmRh$_6$Ge$_4$ first increase rapidly with increasing magnetic field, and then tend toward saturation. The clear hysteresis loops further confirm the FM nature of the magnetic ordering. It should be noted that even if the magnetic field is along easy magnetization direction, the saturation moment of SmRh$_6$Ge$_4$ is only $\sim$ 0.34 $\mu_{\rm B}$, which could be related to a mixed Sm-valence \cite{Mazumda1994,NapGM1978} or a complex hybridization process of multiple $4f$ electrons in Sm ions \cite{WOS:000454615100018}.

\begin{figure}[htbp]
	\begin{center}
		\includegraphics[width=0.95\columnwidth]{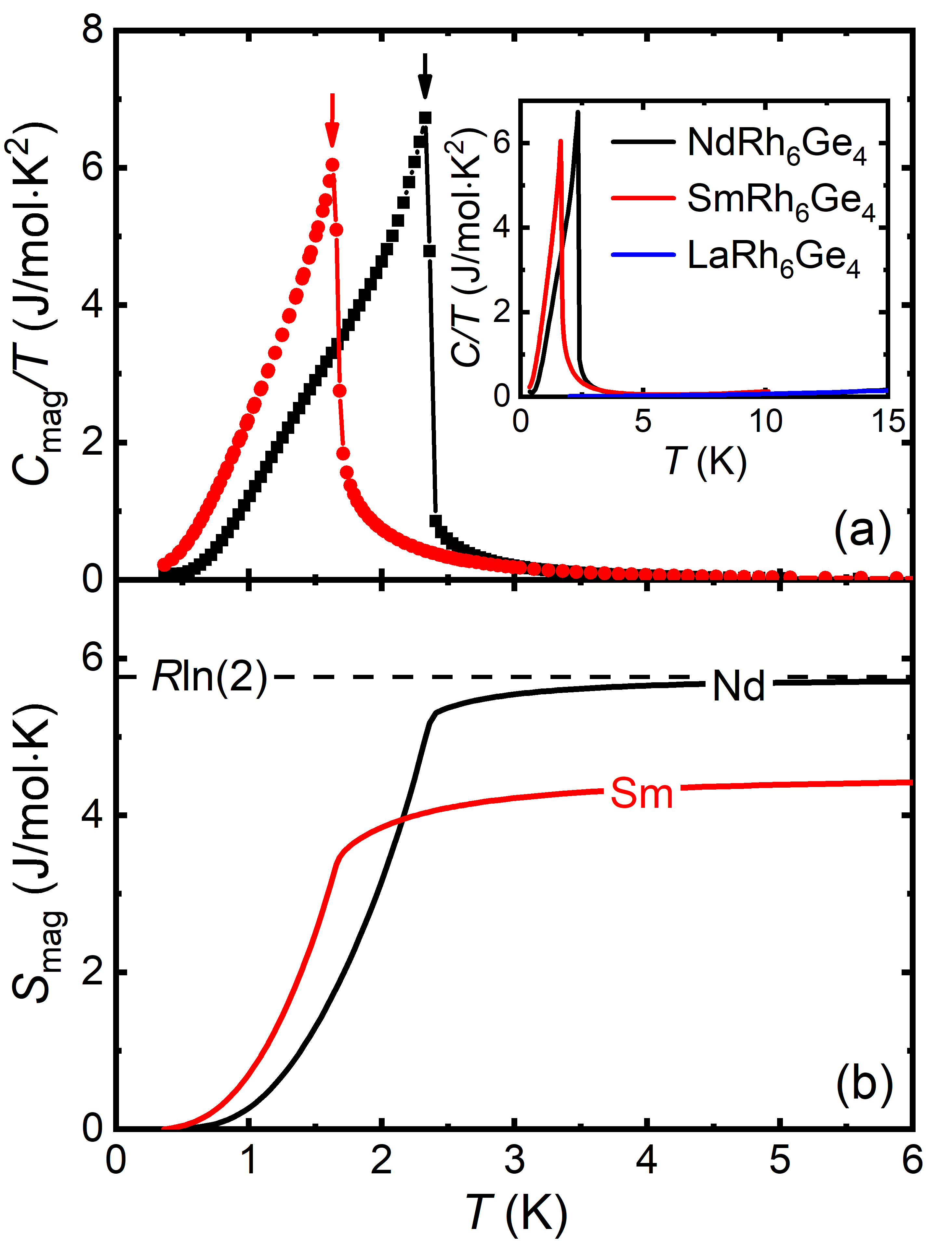}
	\end{center}
	\caption{(Color online) (a) Magnetic part of the specific heat $C_{\rm mag}/T$ of NdRh$_6$Ge$_4$ (black squares) and SmRh$_6$Ge$_4$ (red circles). The arrows show the magnetic transitions. \red{The inset displays the total specific heat as $C/T$ of NdRh$_6$Ge$_4$ (black line), SmRh$_6$Ge$_4$ (red line) and LaRh$_6$Ge$_4$ (blue line).} (b) Temperature dependence of the magnetic entropy $S_{\rm mag}$.}
	\label{figure5}
\end{figure}

The low-temperature magnetic specific heat $C_{\rm mag}/T$ for NdRh$_6$Ge$_4$ and SmRh$_6$Ge$_4$ are shown in Fig. \ref{figure5}(a), which are obtained by subtracting the data of nonmagnetic LaRh$_6$Ge$_4$. \red{The total specific heat of NdRh$_6$Ge$_4$, SmRh$_6$Ge$_4$ and LaRh$_6$Ge$_4$ are displayed in the inset.} Both compounds exhibit a typical $\lambda$-shape peak at $T_{\rm C}$, corresponding well to the transitions observed in $\rho(\rm T)$ and $\chi(\rm T)$. 
By integrating $C_{\rm mag}/T$ for each compound, the magnetic entropy $S_{\rm mag}$ is shown in Fig. \ref{figure5}(b). For NdRh$_6$Ge$_4$, $S_{\rm mag}$ at $T_{\rm C}$ is close to $R{\rm ln}2$, indicating a well separated ground state Kramers doublet. Meanwhile the $S_{\rm mag}$ of SmRh$_6$Ge$_4$ only reaches 75\% of $R{\rm ln}2$ at $T_{\rm C}$. The ground state of SmRh$_6$Ge$_4$ is expected to also be a Kramers doublet and the loss of magnetic entropy could be due to mixed valence or hybridization.

\red{It is worth note that SmRh$_6$Ge$_4$ shares some similar properties with CeRh$_6$Ge$_4$, such as both being easy-plane ferromagnets with a low $T_{\rm C}$, where the saturation moment and magnetic entropy at $T_{\rm C}$ are also lower than the expected values.}

\subsection{D. Antiferromagnetic GdRh$_6$Ge$_4$}

\begin{figure}[htbp]
	\begin{center}
		\includegraphics[width=\columnwidth]{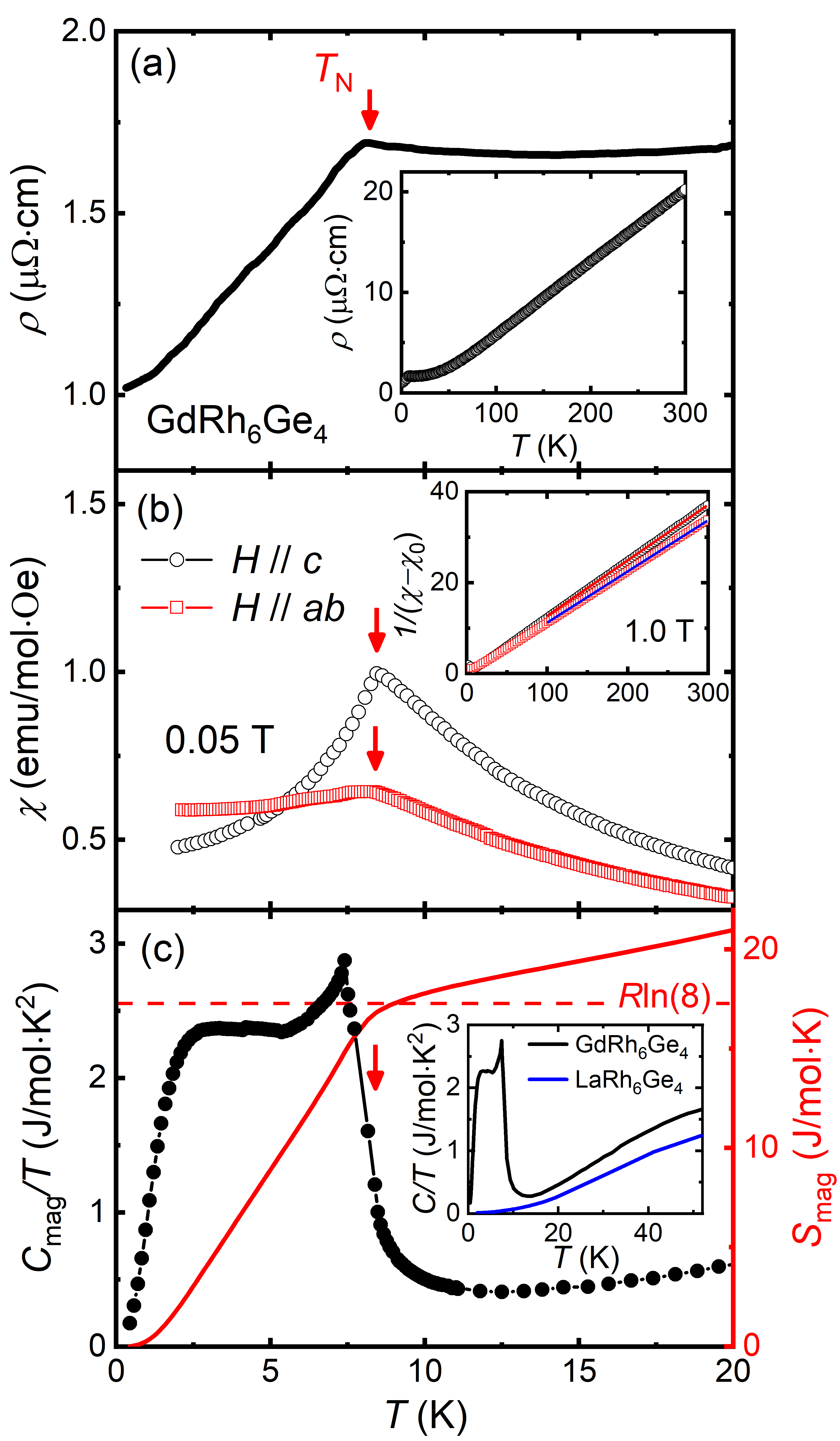}
	\end{center}
	\caption{(Color online) (a) Temperature dependence of the resistivity $\rho(\rm T)$ of GdRh$_6$Ge$_4$ in the temperature range 0.5 - 20 K. The inset displays $\rho(\rm T)$ for 0.5 - 300 K. (b) Temperature dependence of the magnetic susceptibility $\chi(\rm T)$ of GdRh$_6$Ge$_4$ in the temperature range 2 - 20 K, measured in $\mu_0H$ = 0.05 T for $H$ $\Vert$ $c$ (circles) and $H$ $\Vert$ $ab$ (squares). The inset shows the inverse magnetic susceptibility $1/\chi_{\rm M}$. (c) $C_{\rm mag}/T$ (left axis) and $S_{\rm mag}$ (right axis) of GdRh$_6$Ge$_4$. \red{The inset displays the total specific heat as $C/T$ of GdRh$_6$Ge$_4$ (black line) and LaRh$_6$Ge$_4$ (blue line). The red arrows indicate the magnetic transition.}}
	\label{figure6}
\end{figure}

The $\rho(\rm T)$ of GdRh$_6$Ge$_4$ is displayed in Fig. \ref{figure6}(a) . At low temperatures, $\rho(\rm T)$ shows a peak at $T_{\rm N}$ $\sim$ 8.4 K, consistent with the previous report \cite{VosswinkelD2013}. The $\rho(\rm T)$ in the whole temperature range is plotted in the inset of Fig. \ref{figure6}(a), showing typical metallic behavior.

Figure \ref{figure6}(b) displays the temperature dependence of $\chi(\rm T)$ of GdRh$_6$Ge$_4$ measured in $\mu_0H$ = 0.05 T for $H$ $\Vert$ $c$ and $H$ $\Vert$ $ab$. The low temperature $\chi(\rm T)$ for both field directions exhibit sharp cusps at $T_{\rm N}$, indicative of an AFM transition.
As shown in the inset of Fig. \ref{figure6}(b), the inverse magnetic susceptibility $1/\chi$ for $H$ $\Vert$ $c$ and $H$ $\Vert$ $ab$ almost coincide, and both exhibit a linear behavior above 100 K. Based on the modified Curie-Weiss fit, \red{$\chi_0$ = -0.0041(5) emu/mol$\cdot$Oe was obtained for $H$ $\Vert$ $ab$, while $\chi_0$ = -0.00044(6) emu/mol$\cdot$Oe for $H$ $\Vert$ $c$}. The effective magnetic moments of GdRh$_6$Ge$_4$ are $\mu_{\rm eff}^{ab}$ = 8.38 $\mu_{\rm B}$/Gd and $\mu_{\rm eff}^{c}$ = 8.03 $\mu_{\rm B}$/Gd, respectively, which are very close to the theoretical Gd$^{3+}$ free ion value of 7.9 $\mu_{\rm B}$/Gd. The values of $\theta_{\rm p}$ are 1.64 K for $H$ $\Vert$ $ab$ and 0.48 K for $H$ $\Vert$ $c$, which are both relatively small compared to $T_{\rm N}$. 

The magnetic specific heat $C_{\rm mag}/T$ and magnetic entropy $S_{\rm mag}$ of GdRh$_6$Ge$_4$ are displayed in Fig. \ref{figure6}(c). \red{The total specific heat of GdRh$_6$Ge$_4$ and LaRh$_6$Ge$_4$ are displayed in the inset.} A clear sharp peak is observed at $T_{\rm N}$ = 8.4 K, in line with that of $\rho(\rm T)$ and $\chi(\rm T)$.
In addition, there is a wide bulge below $T_{\rm N}$, which likely arises from the Schottky effect common for Gd or Eu compounds with a highly degenerate 4$f$ ground state multiplet.
The $S_{\rm mag}$ of GdRh$_6$Ge$_4$ reaches a value close to $R{\rm ln}8$ at $T_{\rm N}$, which is consistent with the expected isotropic 4$f$-orbitals of the fully degenerate ground state multiplet for the Gd$^{3+}$ ions with $S$ = 7/2 and $L$ = 0.

\begin{figure}[htbp]
	\begin{center}
		\includegraphics[width=\columnwidth]{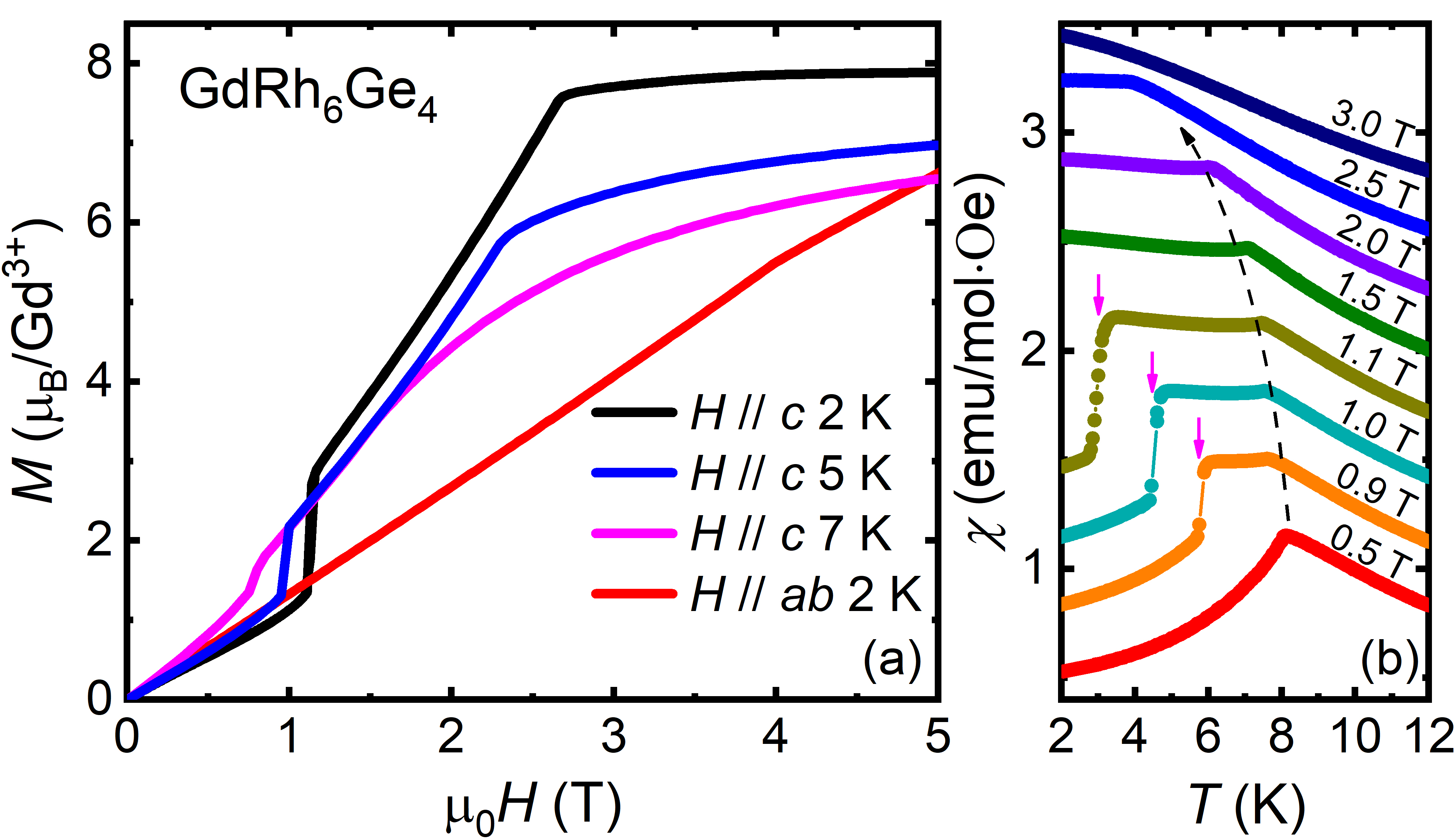}
	\end{center}
	\caption{(Color online) \red{(a) Field dependence of the magnetization of GdRh$_6$Ge$_4$ measured at 2 K (black line), 5 K (blue line), 7 K (magenta line) for $H$ $\Vert$ $c$ and 2 K (red line) for $H$ $\Vert$ $ab$. (b) Temperature dependence of $\chi(\rm T)$ measured in fields applied parallel to the c axis. $\chi(\rm T)$ in consecutive fields are shifted vertically by 0.3 emu/mol$\cdot$Oe for clarity. The dashed arrow denotes the trend of $T_{\rm N}$, while the magenta arrows mark the transition to the field-induced phase.}}
	\label{figure7}
\end{figure}

The field dependence of the magnetization of GdRh$_6$Ge$_4$ measured at several temperatures is displayed in Fig. \ref{figure7}(a), for fields applied along the $c$-axis and in the $ab$-plane.
\red{At 2 K,} upon increasing the $c$-axis magnetic field, there is a metamagnetic transition at $\sim$ 1.1 T with an abrupt jump in the magnetization, which is absent in polycrystalline samples \cite{VosswinkelD2013}. Subsequently, $M(\rm H)$ continues to increase almost linearly, and saturates above 2.7 T. The metamagnetic transition moves to lower field as increasing the temperature.
Whereas for $H$ $\Vert$ $ab$ the $M(\rm H)$ increases monotonically with increasing magnetic field, and there is no obvious metamagnetic transition up to 5 T, although there is a slight change of the slope at 4 T.

\red{Figure \ref{figure7}(b) shows the magnetic susceptibility of GdRh$_6$Ge$_4$ for different magnetic fields with $H$ $\Vert$ $c$. With increasing field, $T_{\rm N}$ is suppressed to lower temperatures (indicated by the black dashed arrow).
While above 0.9 T, a new transition appears in the magnetically ordered state, below which there is a sharp increase in $\chi(\rm T)$, which also shifts to lower temperature with increasing field (magenta arrows).}

\subsection{E. Magnetization plateaus in $R$Rh$_6$Ge$_4$ ($R$ = Tb, Dy, Ho)}

\begin{figure}[htbp]
	\begin{center}
		\includegraphics[width=0.95\columnwidth]{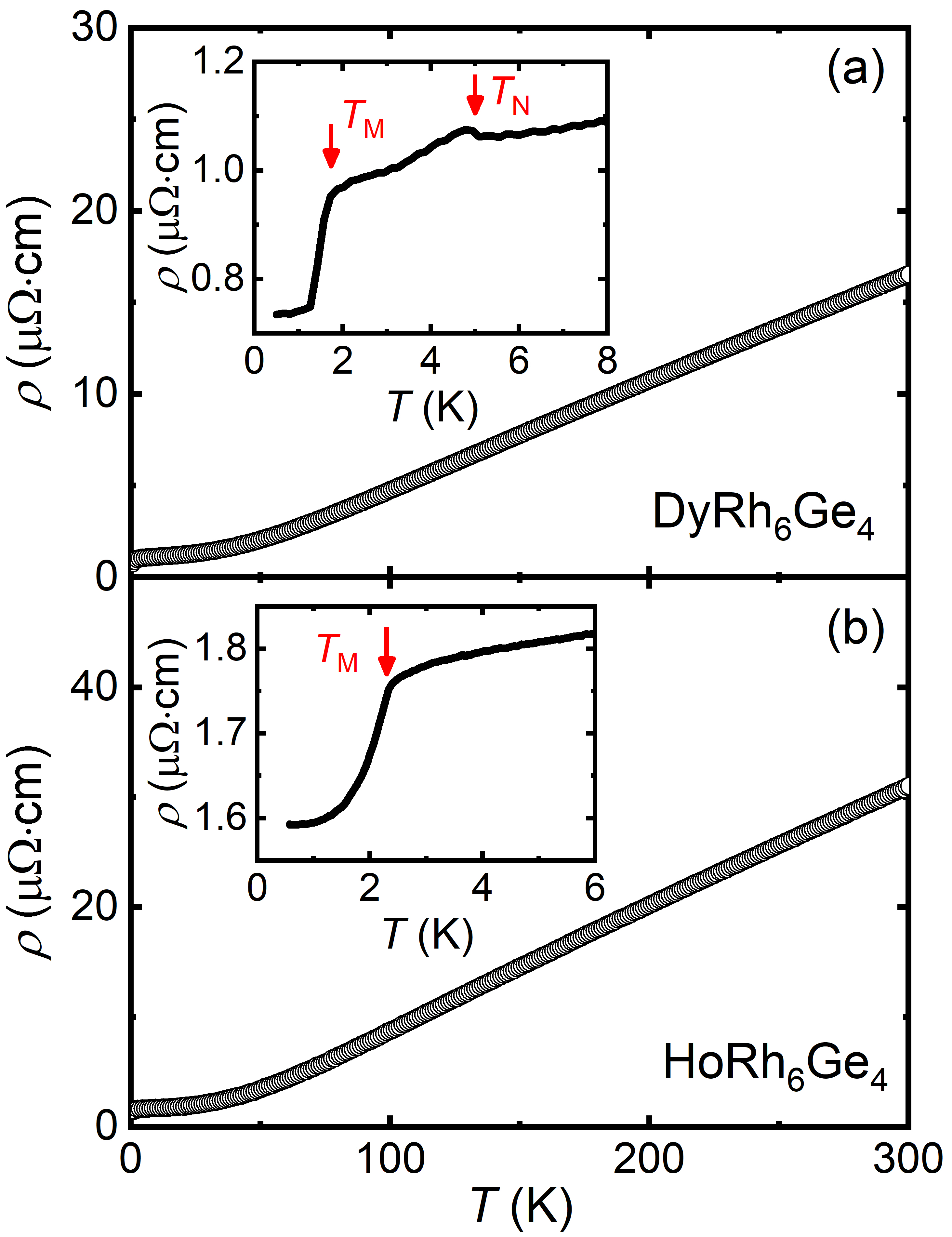}
	\end{center}
	\caption{(Color online) Temperature dependence of the resistivity $\rho(\rm T)$ of (a) DyRh$_6$Ge$_4$ and (b) HoRh$_6$Ge$_4$. Insets show $\rho(\rm T)$ at low temperatures, where red arrows indicate the magnetic ordering temperatures.}
	\label{figure8}
\end{figure}

Figure \ref{figure8} displays the temperature dependence of $\rho(\rm T)$ of DyRh$_6$Ge$_4$ and HoRh$_6$Ge$_4$. At high temperatures, $\rho(\rm T)$ follows typical metallic behavior. At low temperatures, $\rho(T)$ of DyRh$_6$Ge$_4$ shows a peak at the AFM ordering temperature $T_{\rm N}$ $\sim$ 5 K, consistent with the previous report \cite{VosswinkelD2013}, and there is another obvious transition at lower temperature at $T_{\rm M}$ $\sim$ 1.7 K. The $\rho(T)$ of HoRh$_6$Ge$_4$ shows sharp drops at 2.3 K.

\begin{figure*}[htbp]
	\begin{center}
		\includegraphics[width=0.9\textwidth]{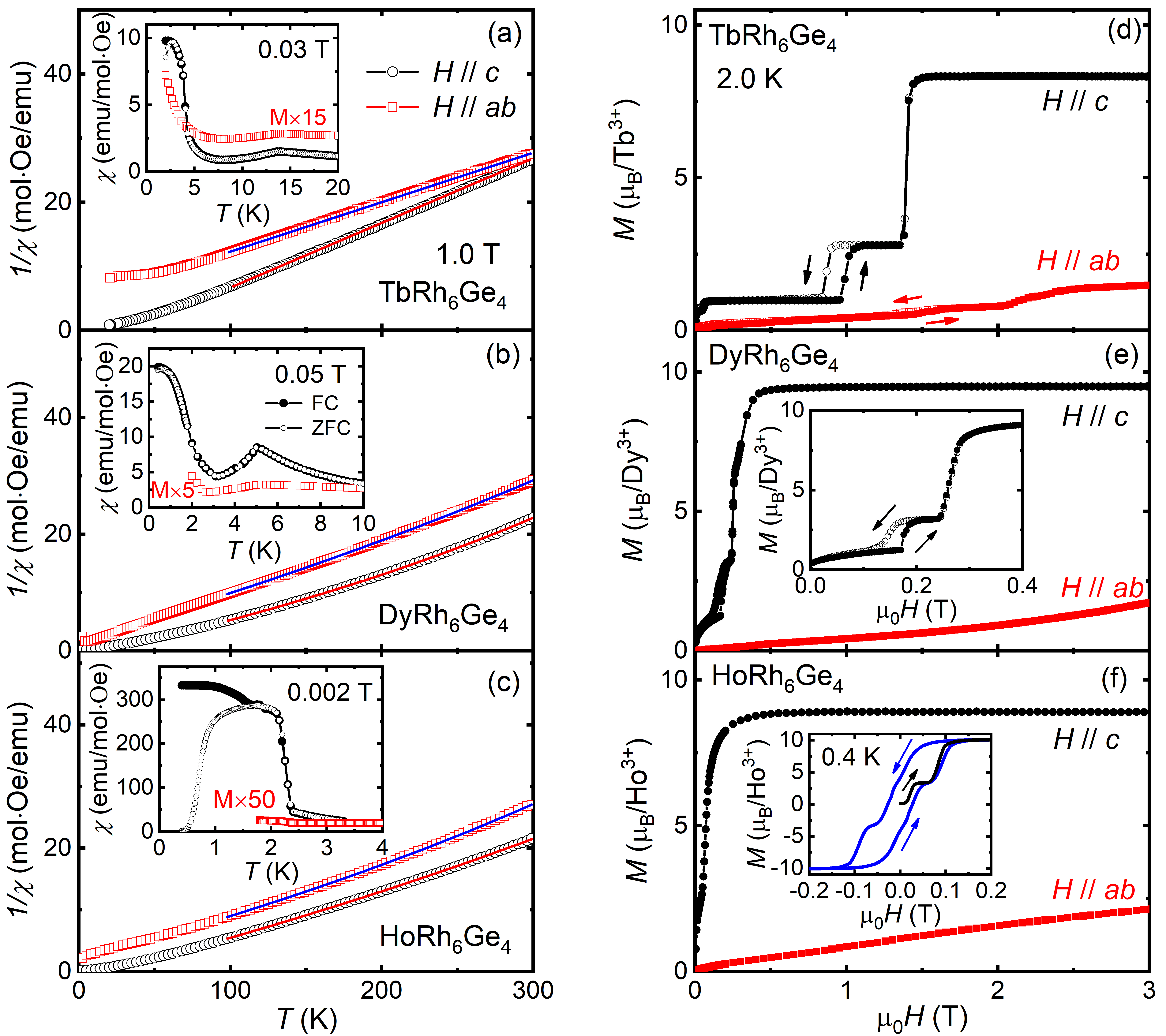}
	\end{center}
	\caption{(Color online) Temperature dependence of the inverse magnetic susceptibility of $R$Rh$_6$Ge$_4$ ($R$ = Tb, Dy, Ho) measured in $\mu_0H$ = 1 T for $H$ $\Vert$ $c$ (circles) and $H$ $\Vert$ $ab$ (squares), where the solid lines correspond to fitting with the modified Curie-Weiss law. Insets show the magnetic susceptibility $\chi(\rm T)$ at low temperatures. \red{Open and closed symbols are data measured upon ZFC and FC, respectively.} (d-f) show the field dependence of the magnetization for $H$ $\Vert$ $c$ (circles) and $H$ $\Vert$ $ab$ (squares) at 2.0 K. Closed and open symbols are data taken while increasing and decreasing magnetic fields, respectively. \red{The inset of (e) is expanded plots in low fields to show the step clearly. The inset of (f) shows the virgin curve (black line) and the magnetization loop with hysteresis (blue line) for $H$ $\Vert$ $c$ at 0.4 K.} The data of TbRh$_6$Ge$_4$ is reproduced from Ref. \cite{ChenY2023} for comparison.}
	\label{figure9}
\end{figure*}

The inverse magnetic susceptibility $1/\chi$ of $R$Rh$_6$Ge$_4$ ($R$ = Tb, Dy, Ho) are plotted in Fig. \ref{figure9}(a) - (c) for $H$ $\Vert$ $c$ and $H$ $\Vert$ $ab$ with $\mu_0H$ = 1 T.
\red{Fitted by the modified Curie-Weiss law, $\chi_0$ of DyRh$_6$Ge$_4$ is -0.0157(4) emu/mol$\cdot$Oe for $H$ $\Vert$ $c$, and -0.0078(2) emu/mol$\cdot$Oe for $H$ $\Vert$ $ab$, while $\chi_0$ of HoRh$_6$Ge$_4$ is -0.0088(1) emu/mol$\cdot$Oe for $H$ $\Vert$ $c$, and -0.0113(2) mu/mol$\cdot$Oe for $H$ $\Vert$ $ab$, respectively.}
Effective moments $\mu_{{\rm eff}}$ and the Weiss temperatures $\theta_{\rm p}$ for different field directions are summarized in Table \ref{table2}.
The estimated effective moments of $R$Rh$_6$Ge$_4$ ($R$ = Tb, Dy, Ho) are close to those of the free $R^{3+}$-ion values and therefore the magnetic properties of these compounds can be explained by considering $4f$ moments of the rare-earth ions. $R$Rh$_6$Ge$_4$ ($R$ = Tb, Dy, Ho) exhibit anisotropic $\theta_{\rm p}$, where $\theta_{\rm p}$ is negative within the $ab$-plane and positive along the $c$-axis.
The insets of Fig. \ref{figure9}(a) - (c) show the magnetic susceptibility $\chi(\rm T)$ of $R$Rh$_6$Ge$_4$ ($R$ = Tb, Dy, Ho) at low temperatures, all of which evidence a significant magnetic anisotropy. The easy magnetization direction is along the $c$-axis for these compounds.
\red{For TbRh$_6$Ge$_4$ and DyRh$_6$Ge$_4$, there are sharp cusps at 12.7 K and 5 K, respectively, which suggests AFM ordering. At lower temperatures, the $\chi(\rm T)$ of both compounds exhibit a considerable increase around $T_{\rm M}$ and begin to saturate at the lowest measured temperatures for $H$ $\Vert$ $c$, indicating that below $T_{\rm M}$ there could be the onset of a FM component.
The $\chi$(T) of HoRh$_6$Ge$_4$ in the $c$-axis shows a sizable increase below $T_{\rm M}$ $\sim$ 2.3 K and there is a clear splitting between zero-field-cooled (ZFC) and FC measurements, which could be the result of FM-type domains.}

\begin{figure*}[htbp]
	\begin{center}
		\includegraphics[width=0.95\textwidth]{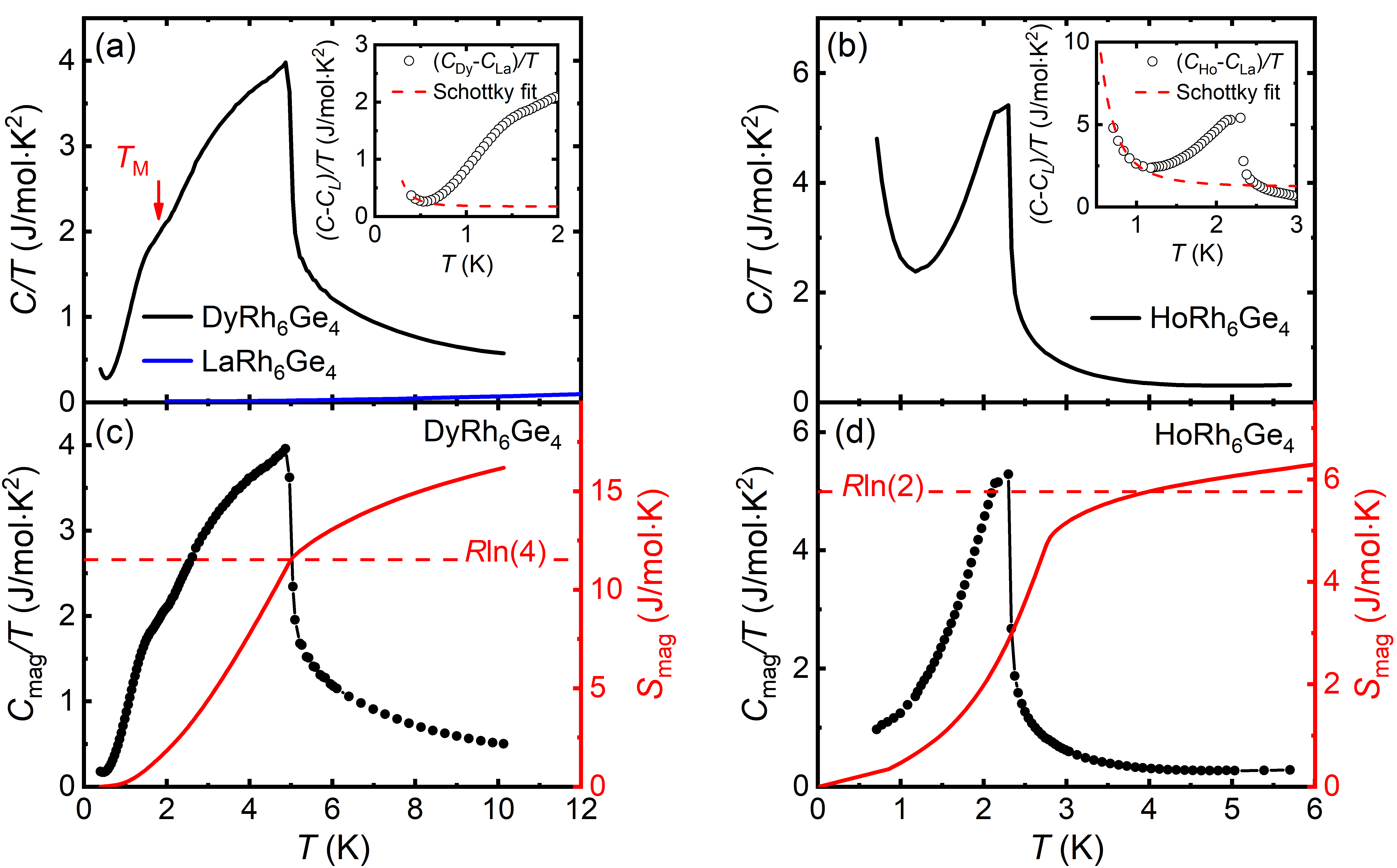}
	\end{center}
	\caption{(Color online) \red{Temperature dependence of the specific heat as $C/T$ of (a) DyRh$_6$Ge$_4$ (black line), (b) HoRh$_6$Ge$_4$ (black line) and LaRh$_6$Ge$_4$ (blue line). The insets show the low-temperature behavior, where the red dashed lines show the results from fitting to a nuclear Schottky contribution at low temperatures. $C_{\rm mag}/T$ (left axis) and $S_{\rm mag}$ (right axis) of (c) DyRh$_6$Ge$_4$ and (d) HoRh$_6$Ge$_4$, where $C_{\rm mag}/T$ is obtained from subtracting both the estimated lattice and nuclear Schottky parts.}}
	\label{figure10}
\end{figure*}

The field dependence of the magnetization, $M(\rm H)$, at $T$ = 2 K for $R$Rh$_6$Ge$_4$ ($R$ = Tb, Dy, Ho) shows a large magnetic anisotropy between $H$ $\Vert$ $c$ and $H$ $\Vert$ $ab$, as shown in Fig. \ref{figure9}(d) - (f). 
\red{Magnetization plateaus at $M_s$/9 and $M_s$/3 are clearly seen in TbRh$_6$Ge$_4$ for $H$ $\Vert$ $c$, where $M_s$ is the saturation magnetization \cite{ChenY2023}.}
For DyRh$_6$Ge$_4$, the magnetization along the $c$-axis has two steep increases at fields of $\sim$ 0.16 T and 0.23 T, respectively, the first of which is accompanied by clear hysteresis. Above 0.26 T, the magnetization gradually increases and finally at 0.34 T reaches $M_s$ $\sim$ 9.3 $\mu_{\rm B}$/Dy. The magnetization around 0.16 T and 0.23 T correspond to $M_s$/9 and $M_s$/3, respectively.
For $H$ $\Vert$ $ab$, the magnetization increases slowly, but the slope of the curve changes slightly around 1.8 T.
\red{As shown in the inset of Fig. \ref{figure9}(f), the virgin curve of HoRh$_6$Ge$_4$ at 0.4 K has two sharp increases at fields of $\sim$ 0.02 T and 0.6 T, respectively, and finally reaches $M_s$ $\sim$ 10.1 $\mu_{\rm B}$/Ho for $H$ $\Vert$ $c$.}
There is a plateau between these increases, the value of which is about $1/3$ of $M_s$. 
\red{A sizable hysteresis is observed in the $M(\rm H)$ at 0.4 K, supporting a ferrimagnetic (FIM) nature of the ground state in HoRh$_6$Ge$_4$. Whereas, the magnetization increases monotonically with the increase of the $ab$-plane magnetic field at 2 K.}

\begin{figure}[htbp]
	\begin{center}
		\includegraphics[width=0.95\columnwidth]{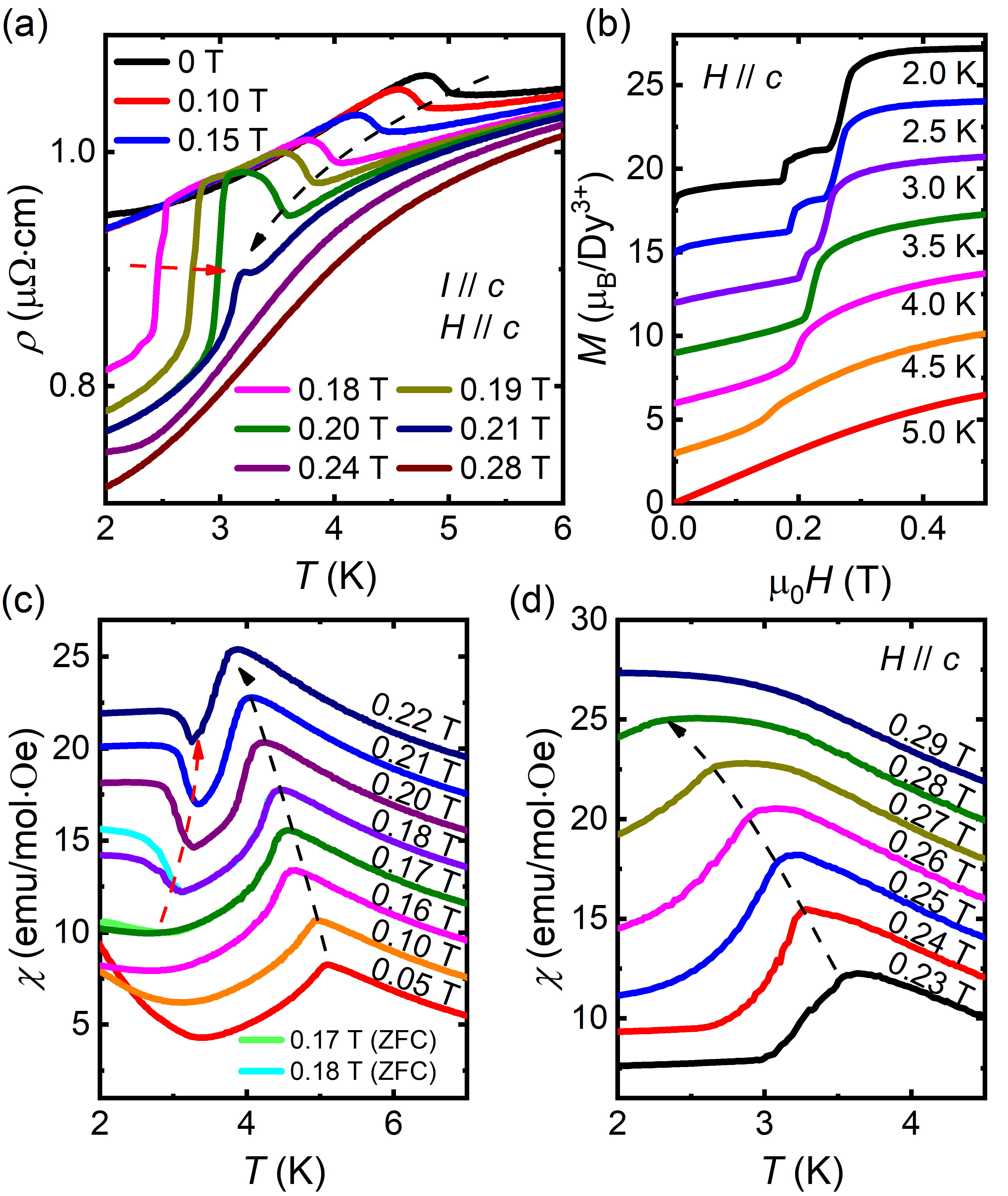}
	\end{center}
	\caption{(Color online) \red{(a) Temperature dependence of the resistivity of DyRh$_6$Ge$_4$ in various fields applied along the $c$ axis, with the current applied along the same direction. The black and red dashed arrows denote the trend of $T_{\rm N}$ and the field-induced transition, respectively. (b) Field dependence of magnetization of DyRh$_6$Ge$_4$ at various temperatures for $H$ $\Vert$ $c$, $M(H)$ at consecutive temperatures are shifted vertically by 3 $\mu_{\rm B}/$Dy for clarity. Temperature dependence of the magnetic susceptibility of DyRh$_6$Ge$_4$ in various fields (c) below 0.22 T and (d) above 0.23 T, applied along the $c$ axis. Data at consecutive fields are shifted vertically by 2 emu/mol$\cdot$Oe for clarity. The black and red dashed arrows denote the trend of $T_{\rm N}$ and the field-induced transition, respectively.}}
	\label{figure11}
\end{figure}

Figure \ref{figure10}(a) and (b) display the temperature dependence of the specific heat as $C/T$ of DyRh$_6$Ge$_4$ and HoRh$_6$Ge$_4$, respectively, in which all exhibit a typical $\lambda$-shape peak at $T_{\rm N}$ ($T_{\rm M}$). In accordance with magnetic susceptibility and resistivity measurements, DyRh$_6$Ge$_4$ shows an additional anomaly at $T_{\rm M}$ $\sim$ 1.8 K. It is worth note that there are upturns below 0.5 K and 1 K for DyRh$_6$Ge$_4$ and HoRh$_6$Ge$_4$, respectively, likely corresponding to a nuclear Schottky contribution that can be described by $C_{\rm N} = A_{\rm N}/T^2$. As shown in the insets of Fig. \ref{figure10}(a) and (b), $A_{\rm N}$ = 0.012 J/mol$\cdot$K$^2$ is obtained for DyRh$_6$Ge$_4$, while $A_{\rm N}$ = 1.37 J/mol$\cdot$K$^2$ for HoRh$_6$Ge$_4$. The magnetic contribution $C_{\rm mag}$ is estimated by subtracting both the lattice contribution $C_{\rm La}$ and $C_{\rm N}$, which are displayed as $C_{\rm mag}/T$, along with the magnetic entropy $S_{\rm mag}$ for DyRh$_6$Ge$_4$ and HoRh$_6$Ge$_4$ in Fig. \ref{figure10}(c) and (d). 
\red{Below 0.7 K, we estimate the magnetic entropy of HoRh$_6$Ge$_4$ by extrapolating the specific heat to zero at 0 K.}
The $S_{\rm mag}$ of HoRh$_6$Ge$_4$ at $T_{\rm N}$ is close to $R{\rm ln}2$, indicating that the ground state of Ho$^{3+}$ is either a doublet or pseudo-doublet. The $S_{\rm mag}$ of DyRh$_6$Ge$_4$ reaches $R{\rm ln}4$ near its $T_{\rm N}$ of 4.8 K, indicating that as well as the ground state Kramers doublet, the magnetic ordering may also incorporate a low-lying excited doublet level.

\begin{figure}[htbp]
	\begin{center}
		\includegraphics[width=\columnwidth]{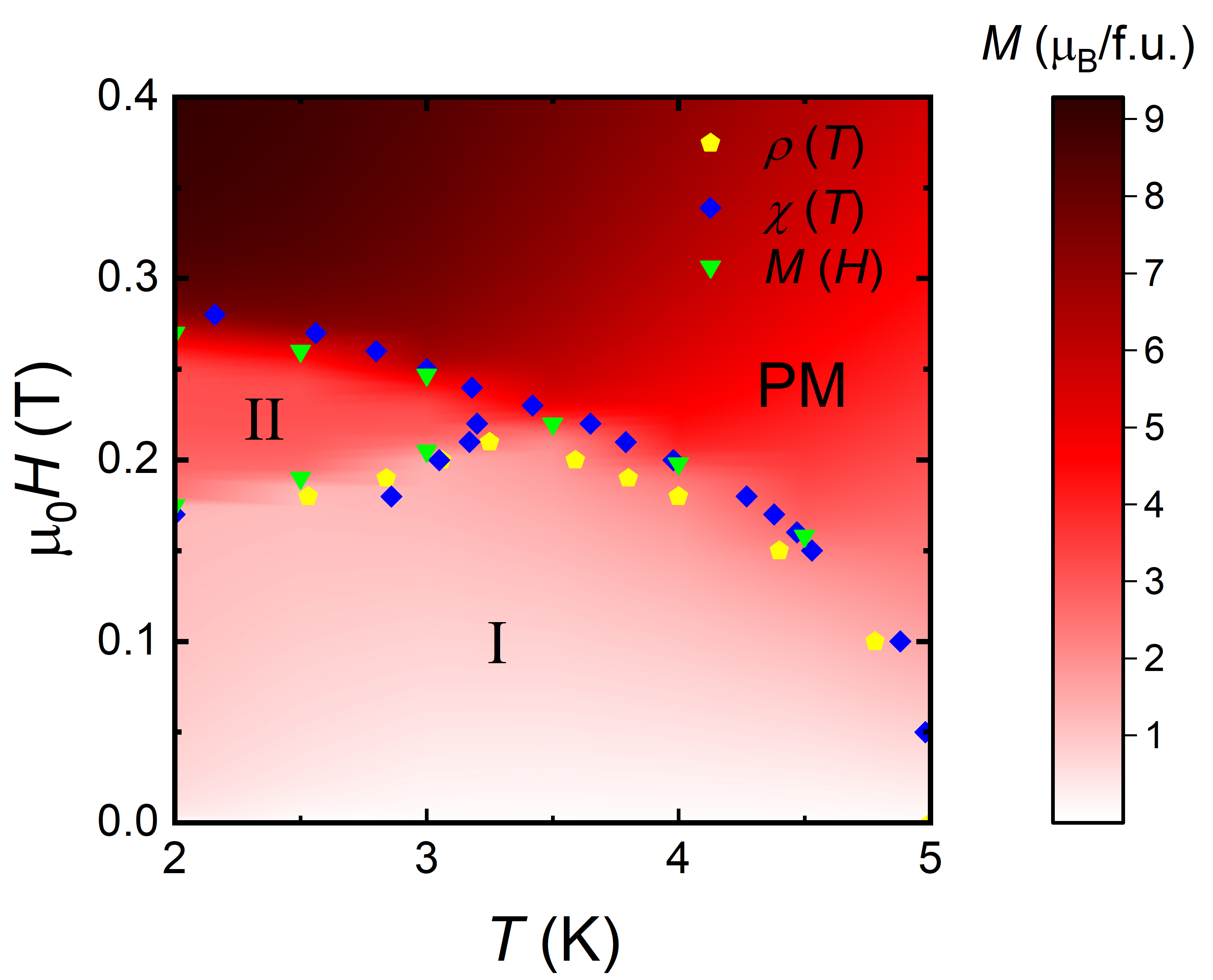}
	\end{center}
	\caption{(Color online) Magnetic field-temperature phase diagram of DyRh$_6$Ge$_4$ for $H$ $\Vert$ $c$. The color plot represents the magnitude of the magnetization, while the symbols correspond to transitions determined from different quantities labeled in the legend. The region labeled $I$ corresponds to the AFM phase below $T_{\rm N}$, while $II$ denotes a field-induced phase.}
	\label{figure12}
\end{figure}

\red{In order to map the $H-T$ phase diagram of DyRh$_6$Ge$_4$, further measurements were performed with different fields or temperatures along the $c$-axis direction. Figure \ref{figure11}(a) shows the low-temperature resistivity of DyRh$_6$Ge$_4$ in various magnetic fields applied along the $c$-axis. With increasing magnetic field, the transition at $T_{\rm N}$ is continuously suppressed to lower temperatures. Upon increasing the field to 0.18 T, a new transition appears at low temperature with an abrupt drop of $\rho(\rm T)$. This field-induced transition shifts to higher temperatures with increasing field and converges with the AFM transition in 0.21 T.}
The isothermal magnetization as a function of field along the $c$-axis at several temperatures is displayed in Fig. \ref{figure11}(b), which was cooled in zero field and then measured upon sweeping the field up. With increasing temperature, the first metamagnetic transition moves to higher field while the second shifts towards lower field. At 3.5 K, there is only a single metamagnetic transition. As the temperature continues to rise, the anomaly moves to lower field, becomes weaker, and vanishes around $T_{\rm N}$ at 5 K.
\red{Figure \ref{figure11}(c) and (d) display the magnetic susceptibility of DyRh$_6$Ge$_4$ for different magnetic fields with $H$ $\Vert$ $c$. With increasing field, $T_{\rm N}$ is suppressed to lower temperatures and the low-temperature increase of $\chi(\rm T)$ seems suppressed upon increasing the field up to 0.16 T. Upon further increasing the field above 0.17 T, a new transition appears in the magnetically ordered state, below which there is a increase in $\chi(\rm T)$, which also shifts to higher temperature with increasing field. Above 2.2 T, only a single transition is observed and becomes broader with further increasing field.}
Finally, we summarize the $H-T$ phase diagram of DyRh$_6$Ge$_4$ ($H$ $\Vert$ $c$) in Fig. \ref{figure12}, and the phase boundaries derived from different measurements coincide well to each other.

\subsection{Discussion}

\begin{table*}[!ht]
	\renewcommand\arraystretch{1.4}
	\centering
	\caption{A summary of the magnetic properties of $R$Rh$_6$Ge$_4$ ($R$ = Ce, Nd, Sm, Gd - Ho): type of magnetism, magnetic ordering temperatures $T_{\rm C}^{\chi}$(K), $T_{\rm N}^{\chi}$(K) and $T_{\rm M}^{\chi}$(K) determined from $\chi$(T); easy magnetization direction; Curie-Weiss temperatures $\theta_{\rm p}$ and effective moment $\mu_{\rm eff}$ for $H$ $\Vert$ $ab$ and $H$ $\Vert$ $c$, respectively; effective moment $\mu_{\rm eff}$ (theoretical value for free $R^{3+}$ ion); and CEF parameter $B_2^0$(K). The data of CeRh$_6$Ge$_4$ and TbRh$_6$Ge$_4$ are from Ref. \cite{PhysRevB.104.L140411} and Ref. \cite{ChenY2023}, respectively for comparison.}
	\begin{tabular}{cccccccccccc}
		\hline
		~$R$~ & ~~type~~ & ~~$T_{\rm C}^{\chi}$(K)~~ & ~~$T_{\rm N}^{\chi}$(K)~~ & ~~$T_{\rm M}^{\chi}$(K)~~ & ~easy-direction~ & ~$\theta_{\rm p}^{ab}$~ & ~$\mu_{\rm eff}^{ab}$~ & ~$\theta_{\rm p}^{c}$~ & ~$\mu_{\rm eff}^{c}$~ &  ~$\mu_{\rm eff}$($\mu_{\rm B}/R^{3+}$)~ & ~$B_2^0$(K)~
		\\ \hline
		Ce & FM & 2.5 &  &  & $ab$-plane & 58 & 1.76 & -133 & 2.19 & 2.53 & 14.5
		\\
		Nd & FM & 2.26 &  &  & $c$-axis & -17.76 & 2.97 & 3.31 & 3.05 & 3.62 & -0.73
		\\
		Sm & FM & 1.65 &  &  & $ab$-plane & - & - & -19.12 & 0.64 & 0.84 & -
		\\
		Gd & AFM &  & 8.4 &  & $c$-axis & 1.64 & 8.38 & 0.48 & 8.03 & 7.9 & -
		\\
		Tb & AFM &  & 12.7 & 2.8 & $c$-axis & -53 & 9.89 & 21.5 & 9.73 & 9.7 & -1.5
		\\
		Dy & AFM &  & 5 & 1.7 & $c$-axis & -25.78 & 10.46 & 18.09 & 11.59 & 10.7 & -0.58
		\\
		Ho & FIM &  &  & 2.28 & $c$-axis & -29.39 & 11.26 & 19.03 & 11.14 & 10.6 & -0.57
		\\ \hline
	\end{tabular}
	\label{table2}
\end{table*}

\begin{figure}[htbp]
	\begin{center}
		\includegraphics[width=\columnwidth]{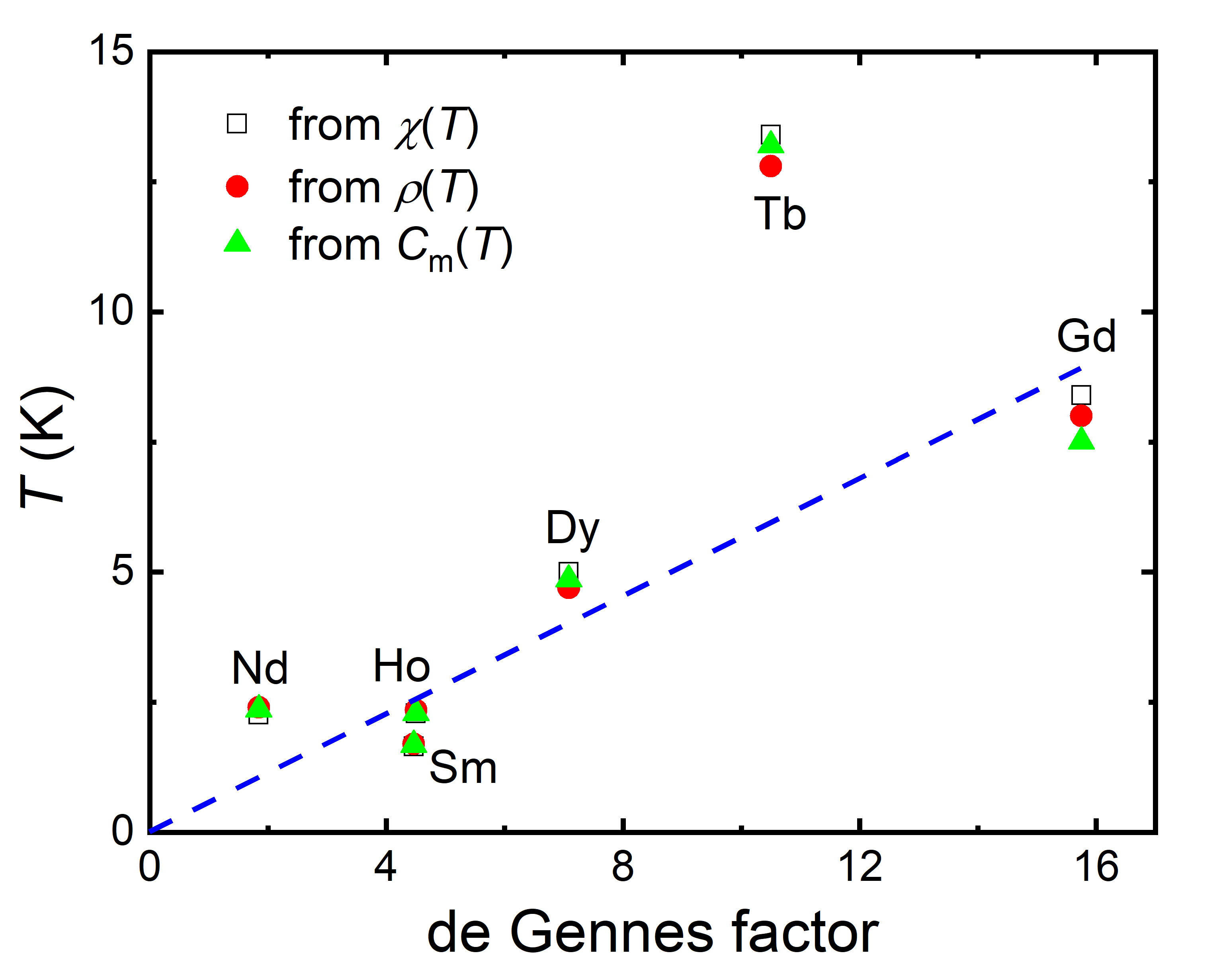}
	\end{center}
	\caption{(Color online) Magnetic ordering temperature, $T_{\rm C}$ or $T_{\rm N}$, as a function of de Gennes factor. The blue dashed line represents expected ordering temperatures for $R$Rh$_6$Ge$_4$ ($R$ = Nd, Sm, Gd - Ho) without CEF.}
	\label{figure13}
\end{figure}

Magnetocrystalline anisotropy plays an important role in governing magnetic properties, which has been revealed in diverse range of systems including $f$-electron materials \cite{JensenJ1991}, transition metal compounds \cite{GerritvanderLaan1998,Daa1991}, and semiconductors \cite{Abo2001,Ale2010}.
The magnetic properties of $R$Rh$_6$Ge$_4$ ($R$ = Ce, Nd, Sm, Gd - Ho) are summarized in Table \ref{table2}.
Figure \ref{figure13} shows the observed magnetic ordering temperatures of $R$Rh$_6$Ge$_4$ as a function of the de Gennes factor. When rare earth ions are the only source of magnetism in intermetallic compounds, the formation of their magnetically ordered states are generally explained by the RKKY exchange interaction. In the molecular field approximation, the magnetic ordering temperature is proportional to the de Gennes factor $(g_{\rm J}-1)^2 J(J+1)$, which is defined as $T_{\rm M}=2/3\mathcal{J}(g_{\rm J}-1)^2 J(J+1)$, where $\mathcal{J}$ is the exchange parameter, $g_{\rm J}$ is the Land\'e g-factor, and $J$ is the total angular momentum quantum number of the ground state of the $R^{3+}$ ion based on Hund's rule \cite{BUD1987,DEG1962}. For rare earth-containing compounds, the magnetic ordering temperature should decrease monotonically as $R$ changes from Gd to Er, and the magnetic ordering temperatures of many rare earth-containing intermetallic compounds follow this scaling \cite{Fal2007,Mar2015,Mor2005,Szy2010}. However, when there are strong CEF effects, for example, $R$ = Tb - Er, it is found that the magnetic ordering temperature of the compound deviates significantly from the linear de Gennes scale \cite{DEG1962,Mar2015}. As shown in Fig. \ref{figure13}, besides TbRh$_6$Ge$_4$, the magnetic ordering temperature of the series of $R$Rh$_6$Ge$_4$ shows only a slight deviation from the linear de Gennes scale. The large deviation of TbRh$_6$Ge$_4$ from the theoretically expected magnetic ordering temperature may be related to its stronger CEF effect.
In addition, in many tetragonal and hexagonal rare-earth-based intermetallic compounds, for $R$ = Tb, Dy, and Ho are easily magnetized in the $c$-axis, while $R$ = Er and Tm favor the $ab$-plane \cite{DEG1962,Bud1999,Mye1999}.
The difference of easy magnetization is due to the change in sign of dominant crystal field parameter. Assuming isotropic magnetic exchange interactions, the CEF parameter $B_2^0$ can be calculated using $B_2^0=10(\theta_{\rm p}^{ab}-\theta_{\rm p}^c)/[3(2J-1)(2J+3)]$ \cite{JensenJ1991}. When there are strong CEF effects, the magnetic ordering temperature depends on the value of $B_2^0$, and a large $B_2^0$ value can increase the magnetic ordering temperature, thus leading to deviations from the simple linear de Gennes scaling \cite{NOAKES198235}.

The estimated $B_2^0$ values of $R$Rh$_6$Ge$_4$ ($R$ = Nd, Tb, Dy, Ho) are also summarized in Table \ref{table2}. Because the high-temperature magnetic susceptibility of SmRh$_6$Ge$_4$ within the $ab$-plane cannot be fitted by the Curie-Weiss law, the $B_2^0$ of SmRh$_6$Ge$_4$ cannot be estimated. Besides this compound and GdRh$_6$Ge$_4$ which lacks CEF effects (since $L$ = 0), all the other compounds have negative $B_2^0$, which coincides with the direction of magnetization being the $c$-axis for these compounds. This indicates that the magnetocrystalline anisotropy is determined by the single-ion CEF effect. 
Following de Gennes scaling, GdRh$_6$Ge$_4$ should have the highest magnetic ordering temperature in this series, but that of TbRh$_6$Ge$_4$ is larger, possibly due to the significant $B_2^0$ value enhancing the ordering temperature. Note that we were unable to synthesize TmRh$_6$Ge$_4$ crystals to investigate its magnetic anisotropy.

It is worth noting that among this series, SmRh$_6$Ge$_4$ and NdRh$_6$Ge$_4$ are found to be ferromagnets with $T_{\rm C}$ of 2.26 K and 1.65 K, respectively. The $B_2^0$ value of NdRh$_6$Ge$_4$ is negative, in contrast to the large positive value in CeRh$_6$Ge$_4$ \cite{PhysRevB.104.L140411}, and for NdRh$_6$Ge$_4$ the easy magnetization direction is along the $c$-axis.
SmRh$_6$Ge$_4$ shares more similar properties with CeRh$_6$Ge$_4$, since both are easy-plane type ferromagnets, which was proposed to be vital for generating the necessary entanglement for allowing the occurrence of a FM QCP \cite{ShenB2020,Komi2018}. It is of high interest to investigate the pressure and doping effects of SmRh$_6$Ge$_4$, which might be another candidate for FM quantum criticality.
Note that if the CEF ground state wave function of SmRh$_6$Ge$_4$ is $\psi^\pm_{GS} = \ket{\pm\frac{1}{2}}$ the same as that of CeRh$_6$Ge$_4$ \cite{PhysRevB.104.L140411}, the estimated $ab$-plane moment $\langle\mu_x\rangle$ = $\bra{\psi^{\mp}}g_J(J^++J^-)/2\ket{\psi^\pm}$ = 0.43 $\mu_{\rm B}$/Sm, which is still larger than the observed low-temperature saturation magnetization of around 0.34 $\mu_{\rm B}$/Sm. 
The reduced value together with the smaller effective high-temperature moment from the Curie-Weiss analysis could be a consequence of the Sm-ions being mixed valence.

In this series, TbRh$_6$Ge$_4$ and DyRh$_6$Ge$_4$ exhibit relatively complex magnetic properties. In zero field both compounds have two magnetic transitions at 12.7 and 2.8 K for TbRh$_6$Ge$_4$, and 5 and 1.7 K for DyRh$_6$Ge$_4$, respectively. Moreover there are multiple metamagnetic transitions when external magnetic fields are applied along the $c$-axis.
The $H-T$ phase diagram of TbRh$_6$Ge$_4$ was previously constructed \cite{ChenY2023}, which shows three magnetic phases and suggests that these metamagnetic transitions correspond to spin-flip transitions of Ising spins that remain strongly constrained to lie along the $c$-axis by magnetocrystalline anisotropy. 
There are plateaus in the $M(\rm H)$ of TbRh$_6$Ge$_4$ where the values correspond to integer fractions of the saturation magnetization $M_s$ of $M_s$/9 and $M_s$/3, respectively. The magnetization plateaus at $M_s$/9 and $M_s$/3 are also observed in DyRh$_6$Ge$_4$. 
Compared to TbRh$_6$Ge$_4$, the field of metamagnetic transitions and the magnetic transition temperature of DyRh$_6$Ge$_4$ are lower, indicating a smaller energy scale of the magnetic exchange interactions. 
For HoRh$_6$Ge$_4$, the magnetization plateau at $M_s$/3 also occurs in the virgin curve of $M(\rm H)$ with a magnetic field along the $c$-axis, \red{although the magnetization loop with hysteresis indicates that its ground state is more likely FIM rather than AFM}. 

The observed multiple magnetization plateaus in $R$Rh$_6$Ge$_4$ ($R$ = Tb, Dy, Ho) along the $c$-axis are also seen in some Ising-like AFM compounds with various structures \cite{DoukoureM1982,BlancoJA1991,gignoux1993magnetic}. For the triangular spin-chain lattice, a two-dimensional (2D) Ising model has been developed to investigate the steplike magnetization \cite{kudasov2006steplike,yao2006steplike}. One typical example is Ca$_3$Co$_2$O$_6$, where the intrachain FM coupling is much stronger than interchain AFM coupling \cite{aasland1997magnetic} and therefore each FM q1D chain can be regarded as a rigid giant spin and then the arrangement is reduced to a 2D triangular lattice composed of giant chain spins, where between the two nearest-neighboring spin chains only the AFM coupling is considered \cite{yao2006steplike}. Using this Ising Hamiltonian, the magnetization plateau at $M_s$/3 is found in agreement with experimental results, which ascribed to two thirds of the chains having spin up and one third having spin down. \red{Compared with Ca$_3$Co$_2$O$_6$, HoRh$_6$Ge$_4$ shares a similar magnetization plateau at $M_s$/3 in the curve measured after zero-field cooling and a magnetization loop with hysteresis.}
Other similar examples are DyAlGa \cite{gignoux1993magnetic} and HoAlGa \cite{ball1992field}, where magnetization plateaus at $M_s$/9 and $M_s$/3 are also observed. Since the interchain AFM coupling cannot be negligible, the spins within each Dy/Ho triangle are frustrated. Neutron diffraction measurements revealed that at zero-field the spins of each chain are antiferromagnetically aligned along the $c$-axis and in the phase with $M_s$/3, one-third of the chains become ferromagnetically aligned \cite{gignoux1993magnetic}. 
For $R$Rh$_6$Ge$_4$ ($R$ = Tb, Dy, Ho), understanding the detailed magnetic structures need further measurements such as neutron or x-ray diffraction. Overall, these findings suggest that $R$Rh$_6$Ge$_4$ is a promising platform for studying the interplay of magnetic interactions and the CEF effect in leading to complex magnetism with magnetization plateaus in rare-earth intermetallic compounds.

\section{Summary}

Single crystals of $R$Rh$_6$Ge$_4$ ($R$ = Pr, Nd, Sm, Gd - Er) are successfully synthesized and their physical properties are investigated by magnetization, specific heat, and resistivity measurements. 
Among them, neither PrRh$_6$Ge$_4$ nor ErRh$_6$Ge$_4$ exhibits magnetic transitions above 0.4 K. NdRh$_6$Ge$_4$ and SmRh$_6$Ge$_4$ show FM transitions at 2.26 K, 1.65 K, respectively, while GdRh$_6$Ge$_4$ and DyRh$_6$Ge$_4$ exhibit AFM transitions at 8 K and 5 K, respectively, whereas  HoRh$_6$Ge$_4$ shows a FIM transition at 2.28 K. 
For $R$ = Nd, Sm, Dy and Ho, $R$Rh$_6$Ge$_4$ exhibits strong magnetocrystalline anisotropy due to the CEF effects on rare-earth ions.
The easy magnetization direction is the $ab$-plane for $R$ = Sm and the $c$-axis for $R$ = Nd, Gd, Dy and Ho. \red{Among them, HoRh$_6$Ge$_4$ has a magnetization plateau at $M_s$/3 in its virgin curve.}
\red{Besides}, DyRh$_6$Ge$_4$ shows \red{an additional} transition at 1.7 K and multiple magnetization plateaus at $M_s$/9 and $M_s$/3. The complex magnetic properties suggest that there is magnetic frustration between the Dy chains, which needs further microscopic measurements to probe the magnetic structure and excitations.

\section{acknowledgments}
This work was supported by the National Key R$\&$D Program of China (Grant No. 2022YFA1402200 and No. 2023YFA1406303), the National Natural Science Foundation of China (Grants No. 12034017, No. W2511006, No. 12222410, No. 12174332, No. U23A20580, No. 12350710785, and No. 12204159)

\end{document}